\documentclass[seceq]{ptptex}

\usepackage{graphicx}
\usepackage{url}
\usepackage{amsmath}
\usepackage{bm}
\usepackage{mathrsfs}

\title{
Harmonic moments of the gluon density distribution in AA collisions%
}

\author{
Yasushi \textsc{Nara}%
}

\inst{
Akita International University, Yuwa, Akita-city 010-1292, Japan
}

\abst{
By using Monte-Carlo implementations of $k_T$-factorization formula
with running-coupling BK unintegrated gluon distributions
for nucleus-nucleus collisions,
we compute higher order harmonic moments of the initial density distribution
for both RHIC(Au+Au@200GeV) and LHC(Pb+Pb@2.76TeV) collisions.
We study their sensitivity to the size of the valence parton
distribution in the nucleon.
}

\begin{document}

\maketitle

\section{Introduction}

Spacetime evolution of the hot and dense system created from heavy ion
collisions at Relativistic Heavy Ion Collider (RHIC) or Large Hadron
Collider (LHC) may be followed by hydrodynamical simulations.  Indeed,
the nearly perfect fluid picture appears to explain large elliptic
flow discovered at RHIC and LHC.  Recently, it was reported the
effects of initial condition for hydrodynamics on the higher order
flow harmonics in Au+Au collisions at RHIC~\cite{Adare:2011tg} by
comparing various hydrodynamical models with different initial
conditions~\cite{Alver:2010dn,Schenke:2010rr,Petersen:2010cw}.
Combined analysis of higher order flow harmonics may
provide strong constraints on both the properties of QGP and
the initial higher order
moments of the density distribution of gluons in coordinate space~\cite{Lacey}.
It was pointed out in Ref.~\cite{Sorensen:2011hm} that all
moments have the same magnitude when there is no length
scale for nucleons inside a nucleus, while their magnitude decreases
with the harmonic number if
a length scale is introduced.  In this work, we compute higher order
harmonic moments within Monte-Carlo version of $k_t$ factorization
formulation with running coupling BK equation, and study the effect of
Gaussian width on the harmonic moments.

\section{Theoretical Model}

We will use the Monte-Carlo implementation of $k_t$ formula
with unintegrated gluon distribution function from numerical solutions of
running coupling Balitsky-Kovchegov
(MCrcBK)~\cite{mckt,Albacete:2011fw} equation
for the computation of gluon production in heavy ion collisions.
This model is a extension of the Monte-Carlo KLN (MC-KLN) model~\cite{MCKLN}. 
First, the nucleons in the two incident nuclei
are randomly sampled according to the Woods-Saxon distribution which
simulates the effects of fluctuations of position of hard color sources.
At each grid point, we compute the thickness function $T_{A}(\bm{r}_\perp)$
to obtain the local saturation scale and then compute the gluon density.
Within a Gaussian nucleon approximation,
thickness function (for the large-$x$ valence partons) is given by
\begin{equation}
T_p(r) = \frac{1}{2\pi B}\exp[-r^2/(2B)]\ .
\end{equation}
Note that the mean square radius of the valence parton distribution
corresponds to $\langle r^2\rangle = 2B$.
The probability of nucleon-nucleon collision $P(b)$ at impact parameter $b$
is
\begin{equation}
P(b) = 1 - \exp[-k T_{pp}(b)],\qquad T_{pp}(b) = \int d^2s \, T_p(s)\,
T_p(s-b)\ .
\end{equation}
where (perturbatively) $k$ corresponds to the product of gluon-gluon
cross section
and gluon density squared.
We fix $k$ so that integral over impact parameter
becomes the nucleon-nucleon inelastic cross section $\sigma_{NN}$ at
the given energy:
\begin{equation}
\sigma_{NN}(\sqrt{s}) = \int d^2b \left(
   1-\exp[-k(\sqrt{s}) \, T_{pp}(b)]
    \right) ,
\end{equation}
with $\sigma_{NN}=61.36$ mb for $\sqrt{s_{NN}}=2.76$ TeV. Hence,
$P(b)$ broadens with increasing energy, even as the size $B$ of the
hard valence partons is fixed.

In this work, we vary the width of the Gaussian
within the range $B=0.2,0.3,0.4$ fm$^2$.
Our model applies the $k_t$-factorized formula
in the transverse plane perpendicular to the beam axis locally.
The number distribution of produced gluons is given by
\begin{equation}
  \frac{dN_g}{d^2 r_{\perp}dy} \sim
   \frac{N_c}{N_c^2-1}
    \int
    \frac{d^2p_\perp}{p^2_\perp}
      \int^{p_\perp} d^2k_\perp \;\alpha_s\,
          \phi_A(x_1,\left(\bm{k}_\perp+\bm{p}_\perp\right)^2)\;
       \phi_B(x_2,\left(\bm{k}_\perp-\bm{p}_\perp\right)^2)~.
      \label{eq:ktfac}
\end{equation}
In MCrcBK~\cite{mckt}, $\phi$ is obtained from the Fourier
transform of the numerical results of the running coupling BK (rcBK)
evolution equation~\cite{Albacete:2007yr}. Note that
in~(\ref{eq:ktfac}) the small-$x$ evolution is {\em not} treated
stochastically and that the density of produced gluons fluctuates only
due to fluctuations of the large-$x$ valence charges.

\section{Initial Harmonic moments}

\begin{figure}[htb]
\includegraphics[width=13.2cm]{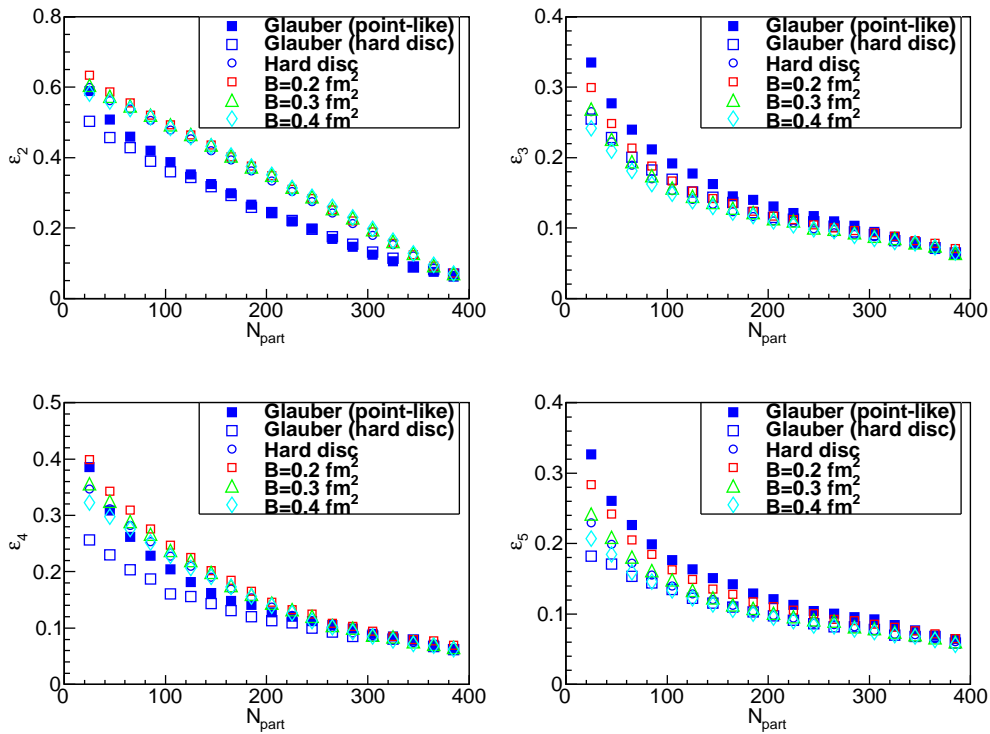}
\caption{Comparison of the centrality dependence of
higher order harmonic moments
for Au+Au collisions at $\sqrt{s_{NN}}=200$ GeV. }
\label{fig:ecc}
\end{figure}

In Fig.~\ref{fig:ecc},
we plot the event average of eccentricities $\varepsilon_n$ ($n=2,3,4,5$)
defined as~\cite{Alver:2010gr}
\begin{equation}
\varepsilon_n = \frac{\sqrt{\langle r^2\cos(n\phi)\rangle^2
+ \langle r^2\sin(n\phi)\rangle^2
}}{\langle r^2 \rangle}
\end{equation}
where $r^2=x^2+y^2$, $x=r\cos\phi$, $y=r\sin\phi$.
$\langle\cdots \rangle$ means the average in the transverse plane.
Since the eccentricities may be proportional to the magnitude of flows,
it is important to know the initial value of these quantities.
The result from Monte-Carlo Glauber model in which eccentricity
is calculated based on the positions of point-like participant
nucleons that is labeled by Glauber (point-like) is shown in
full squares. Monte-Carlo Glauber results assuming hard disc nucleon
are plotted in open squares.
In the hard disc approximation, the thickness function at each grid is
obtained by counting the number of participant nucleons as
$T_A(\bm{r}_\perp) = \frac{\text{number of nucleon within $S$}}{S}$,
where the smearing area $S=42$~mb, at top RHIC energy.
The finite size of the nucleons does not affect 
$\varepsilon_2$ significantly, except for very peripheral collisions.
However, Glauber with point-like nucleons yields larger values for
higher moments ($n\ge3$) which is consistent with the results in
Ref.~\cite{Sorensen:2011hm}.

On the other hand, one observes that the third and fifth harmonics
are similar between the MC-Glauber (hard disc) and the MCrcBK models
except for small value of Gaussian width $B$.
in our model as pointed out in Refs~\cite{Qin:2010pf,Qiu:2011iv}.

We should mention that the prediction of $\varepsilon_3$
from a newly developed event generator DIPSY
based on a dipole model is larger
than that of the MC-KLN model~\cite{Flensburg:2011wx},
while DIPSY prediction for $\varepsilon_2$ is the same as MC-KLN result.
This may due to the additional fluctuations from BFKL cascade
in DIPSY which is not included in our models in this work.

        \begin{figure}[htb]
            \parbox{\halftext}{
    \includegraphics[width=2.7in]{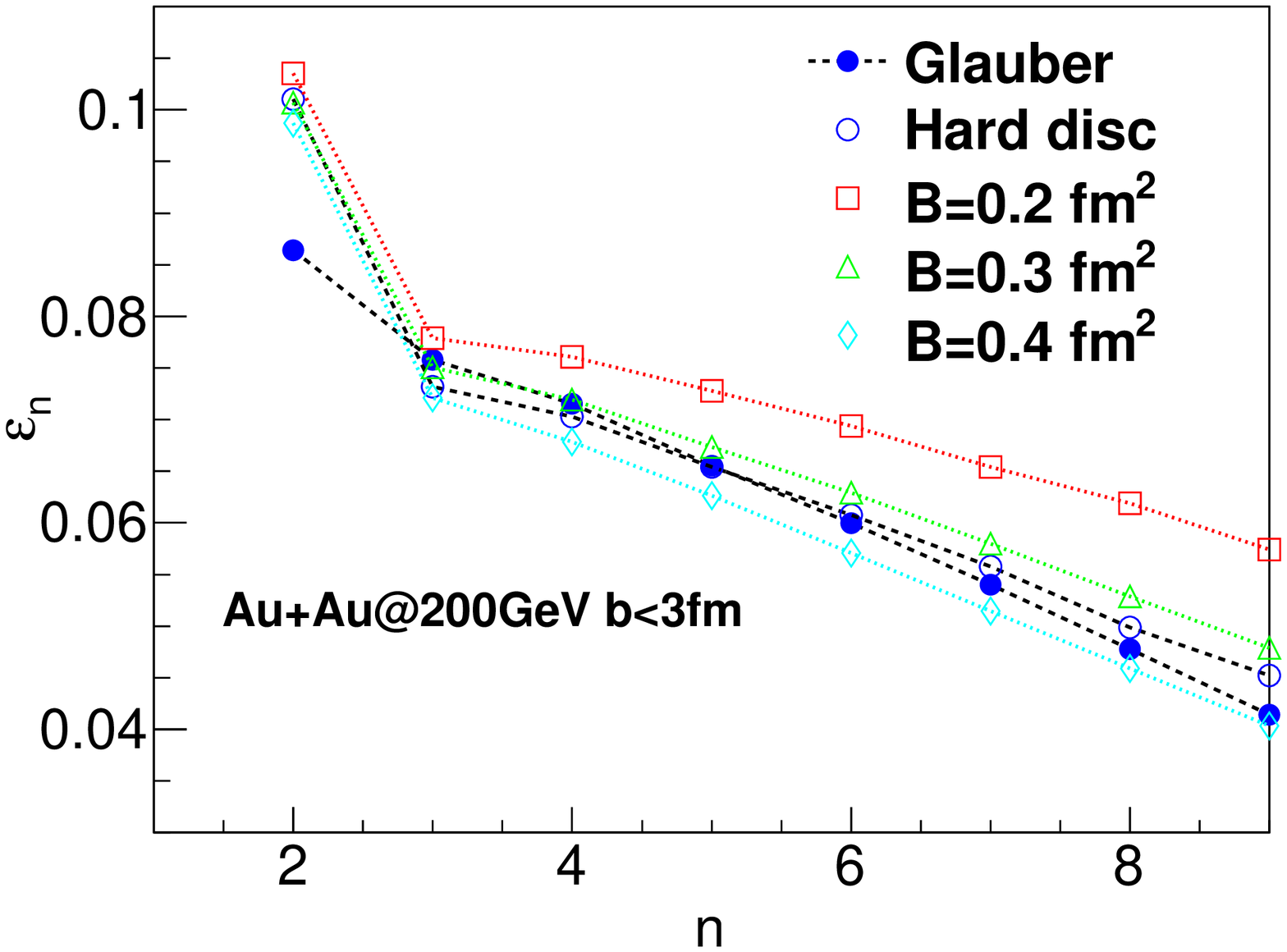}}
            \hfill
            \parbox{\halftext}{
    \includegraphics[width=2.7in]{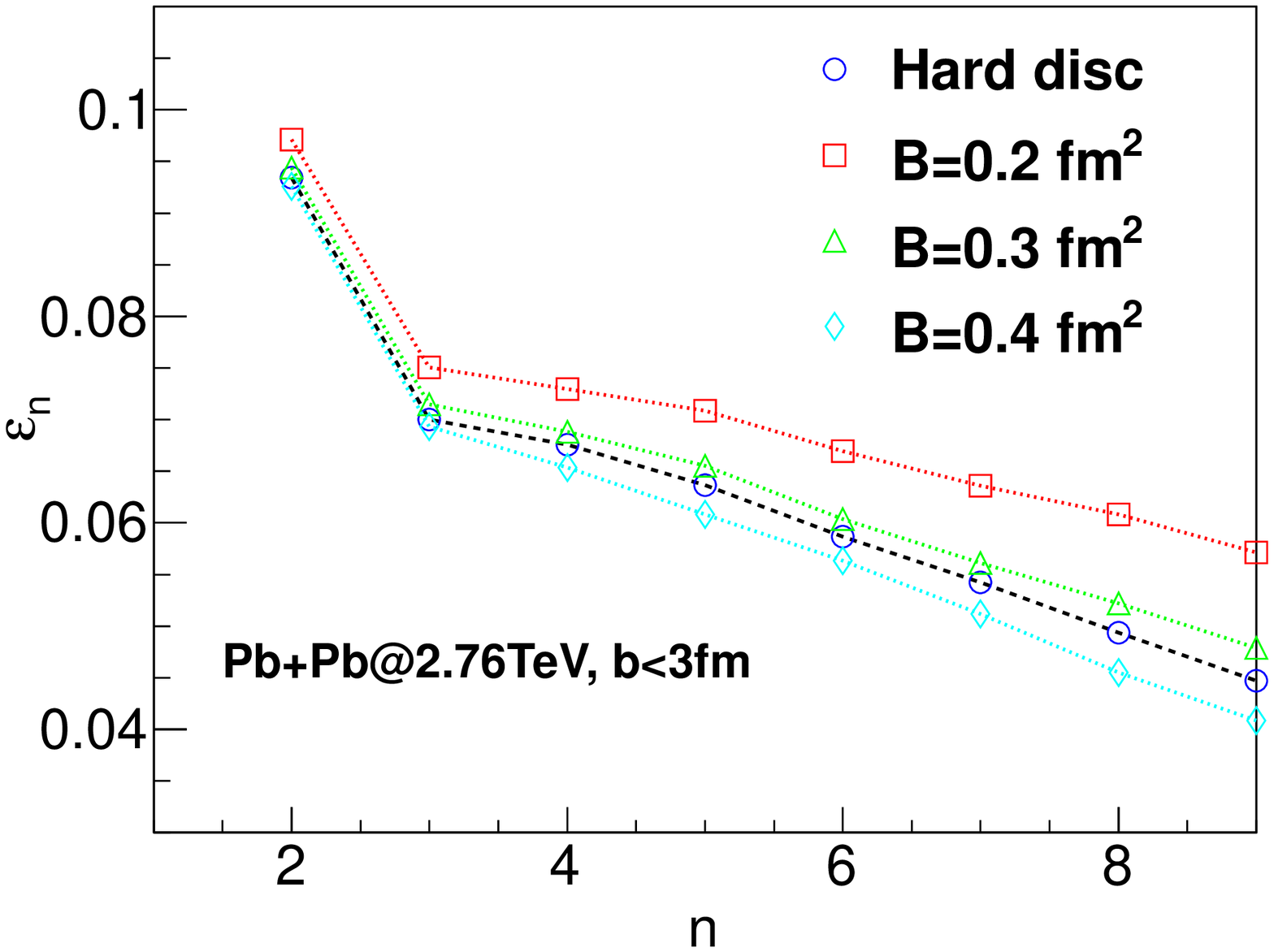}}
                \caption{Harmonic moments for central
		Au+Au collision ($b<3$ fm) at $\sqrt{s_{NN}} = 200$ GeV (left)
                 and
		Pb+Pb collision at $\sqrt{s_{NN}} = 2.76$ TeV (right)
		from MCrcBK with various Gaussian width.}
		\label{fig:eps}
        \end{figure}
Figs.~\ref{fig:eps} show the harmonic moments for
Au+Au collisions at $\sqrt{s_{NN}}=200$ GeV and
Pb+Pb collisions at $\sqrt{s_{NN}}=2.76$ TeV
from the MCrcBK model.
One observes that as Gaussian width decreases, higher order harmonic
moments increase. Therefore, it is important to check the sensitivity
of higher flow harmonics to the 
length scale introduced by valence parton distribution
in order to extract detailed information
on the properties of quark-gluon plasma.

\section{Summary}

We have presented results for higher order harmonic moments
from the Monte Carlo version of $k_t$ factorization formula
with rcBK small-$x$ evolution (MCrcBK). The present simulations
account only for the fluctuations of the valence partons in the
transverse plane.
We have shown the length scale dependence of the harmonic moments of
the initial density distribution.
It will be interesting to see how this affects higher order
hydrodynamical flows. Also, in the future it would be interesting to
systematically explore the effect of additional fluctuations from the
BFKL evolution ladders\cite{Flensburg:2011wx}.

\section*{Acknowledgments}
I thank J.~Albacete, A.~Dumitru and P.~Sorensen for helpful discussions.
This work was partly supported by
Grant-in-Aid for Scientific Research
No.~20540276.

\end{document}